\documentclass[aps,prev,12,twocolumn,preprintnumbers,floatfix,nofootinbib]{revtex4-1}
\pdfoutput=1
\usepackage{graphicx}
\usepackage{bm}
\usepackage{times}
\usepackage{slashed}
\usepackage{color}
\usepackage{color}

\def\ls{\mathrel{\lower4pt\vbox{\lineskip=0pt\baselineskip=0pt
           \hbox{$<$}\hbox{$\sim$}}}}
\def\gs{\mathrel{\lower4pt\vbox{\lineskip=0pt\baselineskip=0pt
           \hbox{$>$}\hbox{$\sim$}}}}
\def\drawbox#1#2{\hrule height#2pt

\hbox{\vrule width#2pt height#1pt \kern#1pt
              \vrule width#2pt}
              \hrule height#2pt}

\def\Asym#1#2{\vcenter{\vbox{\drawbox{#1}{#2}
              \kern-#2pt       
              \drawbox{#1}{#2}}}}

\newcommand{\be}{\begin{equation}}
\newcommand{\ee}{\end{equation}}
\newcommand{\bea}{\begin{eqnarray}}
\newcommand{\eea}{\end{eqnarray}}

\newcommand{\gsim}{\lower.7ex\hbox{$\;\stackrel{\textstyle>}{\sim}\;$}}
\newcommand{\lsim}{\lower.7ex\hbox{$\;\stackrel{\textstyle<}{\sim}\;$}}

\newcommand{\met} {{E\!\!\!\!/_{\rm T}}}

\def\bar {\overline}

\begin{document}

\title{Leptoquarks, Dark Matter, and Anomalous LHC Events}

\author{Farinaldo S. Queiroz$^a$}
\email{fdasilva@ucsc.edu}

\author{Kuver Sinha$^b$}
\email{kusinha@syr.edu}

\author{Alessandro Strumia$^{c,d}$}
\email{Alessandro.Strumia@cern.ch}
\affiliation{$^a$Department of Physics and Santa Cruz Institute for Particle
Physics University of California, Santa Cruz, CA 95064, USA\\
$^b$Department of Physics, Syracuse University, Syracuse, NY 13244, USA\\
$^c$Dipartimento di Fisica dell' Universita di Pisa and INFN, Italy\\
$^d$National Institute of Chemical Physics and Biophysics, Tallinn, Estonia}

\date{\today}

\begin{abstract}

Leptoquarks with mass in the region of $550-650$ GeV are a possible candidate for the recent excess seen by CMS in the $eejj$ and $e \nu jj$ channels. We discuss models where leptoquarks decay primarily to dark matter and jets, thereby giving a branching to charged lepton and jet final states that can match data. The confluence of proton decay constraints, dark matter indirect and direct detection data, and Higgs invisible decay bounds results in a handful of predictive models that will be conclusively verified or excluded in upcoming direct detection experiments. Along the way, we present robust limits on such leptoquark models stemming from the muon magnetic moment using current and projected experiment sensitivities, as well as from $K$ and $B$ meson mixing, and leptonic and semi-leptonic meson decays.



\end{abstract}

\maketitle



\section{Introduction} 

Leptoquarks (LQs) are bosons that couple to a quark and a lepton. While not {\textit essential} elements in addressing today's most outstanding questions (such as the gauge hierarchy problem or the identity dark matter), these particles have a venerable history in particle physics, going back at least to the Pati-Salam model \cite{Pati:1974yy}, $SU(5)$ Grand Unification \cite{Georgi:1974sy}, and technicolor \cite{Farhi:1980xs} \footnote{We refer to \cite{Hewett:1997ce} - \cite{Davidson:1993qk} and references therein for more of the early literature. LQs also appear in a variety of other contexts, of which we cite a few examples \cite{Gershtein:1999gp}.}. 

LQs are strongly interacting, and their decay products include leptons. This fact alone makes them promising search candidates at hadron colliders; indeed, both CMS and ATLAS have placed stringent limits on their masses. 

For second generation LQs, CMS has performed searches with $19.6 {\ \rm fb^{-1} }$ of data focusing on the $\mu \mu jj$ and $\mu \nu jj$ final states, which has resulted in the exclusion of scalar LQs with masses below 1070 (785) GeV for ${\rm Br}_{\mu j} = 1 (0.5)$ \cite{CMSresult2,ATLAS:2012aq}. Similarly, third generation scalar LQs with masses below 740 GeV are excluded at $95\%$ C.L assuming $100\%$ branching fraction for the LQ decay to a $\tau$ lepton and a
bottom quark \cite{Khachatryan:2014ura}, whereas the case of branching ratio of $100\%$ into top quark - $\tau$ lepton pairs is ruled out up to a mass of 634 GeV \cite{CMSlepto}.

Searches for first generation of letptoquarks, on the other hand, have resulted in a mild  evidence for  $\sim 650$~GeV scalar LQs. The CMS collaboration using $19.6 {\ \rm fb^{-1}}$ of integrated luminosity has revealed a $2.4 \sigma$  and $2.6 \sigma$ excess in the $eejj$ and $e \nu jj$ channels respectively, after optimizing cuts for a $650$ GeV LQ \cite{CMSresult}. 

An interesting feature of the recent excess seen in the first generation searches is that the branching into lepton and jets final states cannot be the entire story. LQs are produced dominantly through gluon fusion and quark-antiquark annihilation. The NLO production cross section for a $650$ $(550)$ GeV LQ is $0.0132$ $(0.0431)$ pb, with a PDF uncertainty of $0.00322$ $(0.00893)$ pb. If the renormalisation/factorisation scale is varied between half the LQ mass and double the LQ mass, the cross sections are $0.0113$ $(0.037)$ pb and $0.0149$ $(0.0485)$ pb respectively. On the other hand, the observed cross section times the branching to lepton and jets final states is much smaller than the theoretical prediction. In fact, a recent study \cite{Bai:2014xba} of the excess has concluded that for a $550$ GeV LQ, the best fits are obtained for ${\rm Br}_{ej} = 0.12$ and ${\rm Br}_{\nu j} = 0.15$, while for a $650$ GeV LQ, the branching ratios are ${\rm Br}_{ej} = 0.21$ and ${\rm Br}_{\nu j} = 0.13$. In other words, there is a significant branching of LQs into other states, potentially dark matter.

The excess observed is clearly preliminary, and results from the $14$ TeV run should be watched carefully. Nevertheless, it is worthwhile to explore model building possibilities in case the signal does hold up. A theoretically appealing scenario would arise if LQs are tied to dark matter while obeying current constraints. 

The purpose of this paper is to outline possible scenarios that fulfill these criteria. In particular, we study frameworks where a $550-650$ GeV LQ possibly has significant branching to dark matter, and present the confluence of proton decay constraints, indirect and direct dark matter detection bounds and the Higgs invisible decay limits in such a setting. This turns out to be remarkably predictive and verifiable, and for the benefit of the reader, we summarize our results here. 

Proton decay constraints limit the possible scalar LQ models to just two - those where the LQs are doublets under $SU(2)$. The LQ gauge charges pick out the following dark matter candidates (after requiring dark matter stability) - $(i)$ scalar or fermionic triplets, which are ruled out by Fermi-LAT data in the mass range of interest; $(ii)$ scalar or fermionic singlets, which are within reach of the projected XENON1T experiment in the mass range of interest.

Along the way, we present robust limits stemming from the muon magnetic moment using current and projected experiment sensitivities, as well as coming from $K$ and $B$ meson mixing, and leptonic and semi-leptonic meson decays.

The paper is organized as follows. In Section \ref{secModels}, we write down the possible LQ models and discuss relevant proton decay, muon magnetic moment, and meson mixing and decays bounds. In Section III, we write down our model connecting the LQ and dark matter sectors, whereas in Section IV we work out the dark matter phenomenology. Lastly, we end with our conclusions. 
\section{General Leptoquark Models}
\label{secModels}
In this section, we first present simplified models of scalar LQs and discuss stringent bounds stemming from the muon magnetic moment, collider, flavor and proton decay bounds.  A general classification of renormalizable LQ models can be found in \cite{Buchmuller:1986zs,Davies:1990sc}. The relevant models, following the notation of Ref.\cite{Buchmuller:1986zs} are listed in the Table I.
\begin{table}[h!]
\begin{center}
    \begin{tabular}{ c  c  c}
    \hline \hline

     leptoquark & \ \ \ \ \  leptoquark & \ \ \ \ \ $ SU(3) \times SU(2) \times U(1)$ \ \ \ \ \   \\

      notation   & \ \ \ \ \ couplings & \ \ \ \ \ representation of LQ \ \ \ \ \   \\ \hline \hline
       
   $R_2$ &   $R_2 \bar{Q}e$, $R_2 L \bar{u}$ & $\left(3,2, 7/6\right)$  \\  
  $\widetilde{R}_2$ & $\widetilde{R}_2 L\bar{d}$  & $\left(3,2, 1/6\right)$  \\  
 $S_1$ &  $\ \ S_1 \bar{Q}\bar{L}$, $S_1 \bar{u}\bar{e}$  & $\left(3,1, -1/3\right)$  \\  
   $S_3$ &  $S_3 \bar{Q}\bar{L}$ & $\left(3,3, -1/3\right)$  \\ 
  $\widetilde{S}_1$ &  $\widetilde{S}_1 \bar{d}\,\bar{e}$ & $\left(3,1, -4/3\right)$  \\ 
    \end{tabular}
\end{center}
\caption{\footnotesize{Interaction terms for scalar LQs allowed by symmetries. Note that all the LQ candidates have $L = -1, \, B = 1/3, \, B - L = 4/3$. }}
\label{table1}
\end{table}
For reasons that will be clear later, we will be mostly interested in the models containing $R_2$ and $\widetilde{R}_2$, where the latter will be our primary focus when we discuss the dark matter phenomenology and the connection to the recent CMS excess. The model containing $\widetilde{R}_2$ has a Lagrangian given by
\bea \label{lepto_lag1}
\mathcal{L} &=& -\lambda_d^{ij} \bar{d}_{R}^i \widetilde{R}_2^T\epsilon L_L^j + {\rm h.c.}\ ,
\eea
where $(i,j)$ denote flavor indices. Expanding the ${SU(2)}$ indices yields,
\bea
\mathcal{L} &=& -\lambda_d^{ij} \bar{d}_{\alpha R}^i(V_\alpha e_L^j - Y_\alpha \nu_L^j)  + {\rm h.c.}\ .
\eea
Similarly, the model containing $R_2$ has the following Lagrangian: 
\bea
\mathcal{L} &=& -\lambda_u^{ij} \bar{u}_{R}^i R_2^T\epsilon L_L^j -\lambda_e^{ij} \bar{e}_{R}^i R_2^\dagger Q_L^j + {\rm h.c.}\ ,
\eea
where
\bea
R_2 = \left(
      \begin{array}{c}
        V_\alpha \\
        Y_\alpha \\
      \end{array}
    \right) \ , \ \ \ \epsilon = \left(
                                   \begin{array}{cc}
                                     0 & 1 \\
                                     -1 & 0 \\
                                   \end{array}
                                 \right) \ ,  \ \ \ L_L = \left(
                                                            \begin{array}{c}
                                                              \nu_L \\
                                                              e_L \\
                                                            \end{array}
                                                          \right)\ .
\eea
After expanding the ${SU(2)}$ indices it takes the form,
\bea
\mathcal{L} &=& -\lambda_u^{ij} \bar{u}_{\alpha R}^i(V_\alpha e_L^j - Y_\alpha \nu_L^j) \nonumber\\
&&-\lambda_e^{ij} \bar{e}_{R}^i(V_\alpha^\dagger u_{\alpha L}^j + Y_\alpha^\dagger d_{\alpha L}^j) + {\rm h.c.}\ ,
\eea
where the same notation as before has been used.
Now we will turn our attention to discussing important bounds on LQs. 
\subsection{Proton Decay}
A very important issue to be accounted for in LQ models is baryon number violation, due to the strong bounds on processes like $p \rightarrow \pi^{0} e^{+}$. Only those models that have no proton decay in perturbation theory should be considered. Hence the fields $S_1$, $S_3$, and $\tilde{S}_1$ that allow the respective operators $S_1 Q Q+S_1 u d, S_3 Q Q$, and $\widetilde{S}_1 u u$ are ignored since they induce fast proton decay via tree-level scalar LQ exchange. For the LQs $R_2$ and $\tilde{R}_2$, symmetries disallow couplings of these operators to quark bilinears; hence, there is no proton decay diagram via LQ exchange at dimension four \cite{Arnold:2013cva}. However, there are still dangerous dimension five operators, namely,
\begin{figure}[!h]
\centering
\includegraphics[width=6cm]{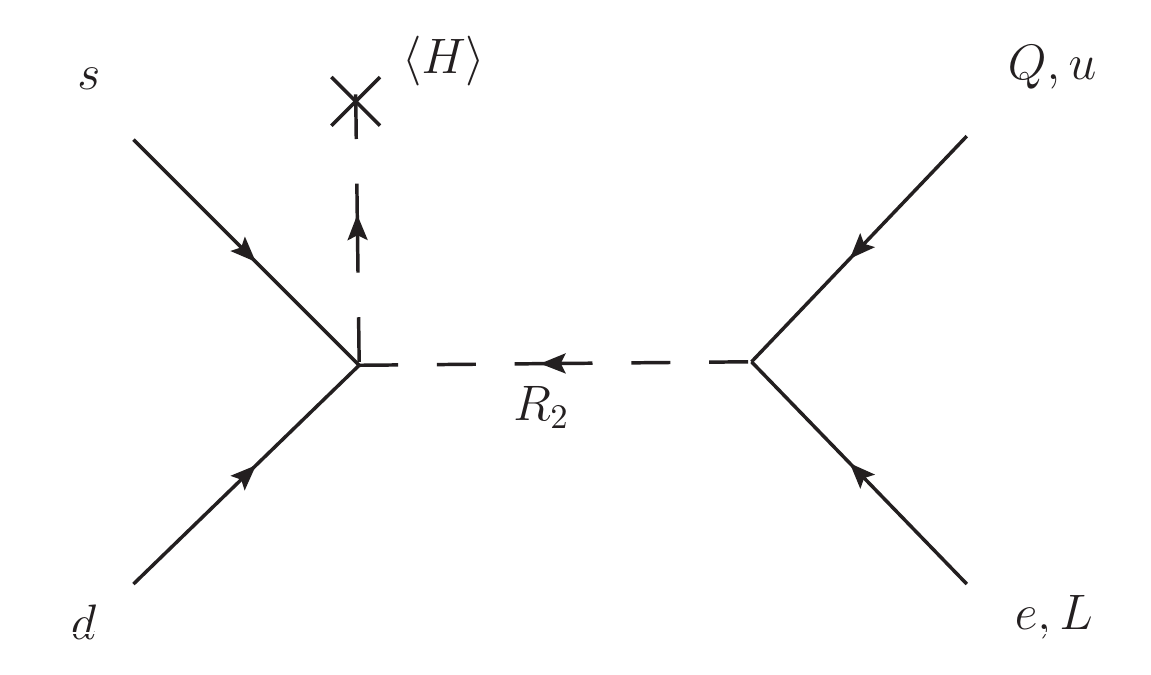}
\caption{Diagram for proton decay.}
\label{protondecay}
\end{figure}
\begin{equation}
\mathcal{L} (R_2)  \supset  \frac{1}{\Lambda} \, g^{a b} d_{R \alpha}^a d_{R \beta}^b (H^\dagger R_{2\gamma}) \epsilon^{\alpha \beta \gamma}
\label{O1}
\end{equation}
\begin{equation}
\mathcal{L} (\widetilde{R}_2)   \supset \frac{1}{\Lambda} \, g^{a b} u_{R \alpha}^a d_{R \beta}^b (H^\dagger \widetilde{R}_{2\gamma}) \epsilon^{\alpha \beta \gamma}
\label{O2}
\end{equation}
\begin{equation}
\mathcal{L} (\widetilde{R}_2)   \supset   \frac{1}{\Lambda} \, g^{a b} u_{R \alpha}^a e_{R}^b (\widetilde{R}_{2\beta}\epsilon \widetilde{R}_{2\gamma}) \epsilon^{\alpha \beta \gamma}\ .
\label{O3}
\end{equation}
For example,  in Fig.~\ref{protondecay} we display a diagram that shows baryon number violating decay involving the LQ $R_2$. We stress that the coupling constant matrix g is antisymmetric in flavor space. Note that Eq.\ref{O1} leads to the nucleon decay $n \rightarrow e^- K^+$ and $p \rightarrow K^+ \nu$, whereas Eq.\ref{O2} induces $n \rightarrow e^- \pi^+$ and $p \rightarrow  \pi^+ \nu$. The lifetime of such decays is short, demanding the LQs masses to lie much above the TeV scale. 

A simple way to prevent these operators is to impose a $Z_3$ symmetry under which fields carry charges  ${\rm exp}[2\pi i(B-L)/3]$ \cite{Arnold:2013cva}. It is easy to check that imposing this symmetry (or any other discrete subgroup of $U(1)_{B-L}$) will not remedy the tree level proton decay induced by the LQs $S_1, S_3,$ and $\tilde{S}_1$, since all the relevant interactions in those cases conserve $B-L$. In conclusion, in order to have a plausible LQ model with proton stability, LQs that are doublets under $SU(2)$ are preferred, and moreover some sort of discrete symmetry must be invoked. Particularly, in the case of LQs $R_2$ and $\widetilde{R}_2$ it is a $Z_3$ one.
\subsection{Muon Magnetic Moment}
The muon magnetic moment is one of the most precisely measured quantities in particle physics, and a $3.6\sigma$ discrepancy has been observed recently indicating that new physics may be around the corner. Despite the uncertainties rising from hadronic contributions, the current deviation is reported as $\Delta a_{\mu} = (295 \pm 81)\times 10^{-11}$ \cite{Queiroz:2014zfa}. The combined effort from the theoretical and experimental side expects to reduce the uncertainties down to $\Delta a_{\mu} = (295 \pm 34)\times 10^{-11}$ \cite{Queiroz:2014zfa}. The latter is referred as our projected bound. In this context, through the diagrams shown in Fig.2, LQs give rise to sizeable contributions according to Eq.(\ref{leptoSmuon1}) \cite{Queiroz:2014zfa},
\begin{eqnarray}
&&
\Delta a_{\mu} (V) = \frac{1}{8\pi^2}\frac{N_c Q_q m_\mu^2}{ M_{V}^2 } \int_0^1 dx \ \frac{g_{s}^2 \ P_{s}(x) + g_{p}^2 \ P_{p}(x) }{(1-x)(1-\lambda^2 x) +\epsilon^2 \lambda^2 x}\nonumber\\+
&&
\frac{1}{8\pi^2}\frac{N_c Q_{\Phi} m_\mu^2}{ M_{\Phi}^2 } \int_0^1 dx \ \frac{g_{s}^2 \ P_{s}^{\prime}(x) + g_{p}^2 \ P_{p}^{\prime}(x) }{\epsilon^2 \lambda^2 (1-x)(1-\epsilon^{-2}x) +x}
\label{leptoSmuon1}
\end{eqnarray}where $\epsilon=m_{q}/m_{\mu}$ and $\lambda=m_{\mu}/M_{V}$, $m_{q} (Q_q)$ is the mass (electric charge) of the quark running in the loop, $g_s$ and $g_a$ are the respective scale and pseudo-scalar couplings to muons, and
\begin{eqnarray}
P_{s}^{\prime}(x) & = &  x^2 (1+\epsilon -x),\nonumber\\
P_{p}^{\prime}(x) & = &  x^2 (1+\epsilon -x),\nonumber\\
P_{s}^{\prime}(x) & = &  -x(1-x)(x+\epsilon),\nonumber\\
P_{p}^{\prime}(x) & = &  -x(1-x)(x-\epsilon)
\label{leptoSmuon2}
\end{eqnarray}

\begin{figure}[!h] \label{muon_gminus2}
\includegraphics[width=3cm]{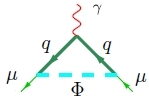}
\includegraphics[width=3cm]{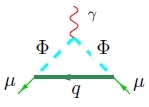}
\caption{Diagrams involving general LQs ($\Phi$) that contribute to muon $g-2$.}
\end{figure}

\begin{table} [htbp]
\begin{center}
{\color{blue} g-2 Bounds on $650$~GeV Leptoquarks}\\
\begin{tabular}{|c||c|}
\hline 
Model  & Limit\\ 
\hline
$R_2$: $1^o$ {\rm Generation} & ${\rm coupling} < 1$ \\
\hline
$R_2$: $2^o$ {\rm Generation} &  ${\rm coupling} < 0.1$ \\
\hline
$R_2$: $3^o$ {\rm Generation} & ${\rm coupling} < 0.01$ \\
\hline
$\tilde{R}_2$: $1^o,2^o,3^o$ {\rm Generation} & No bound \\
\hline
\end{tabular}
\end{center}
\caption{Limits on the couplings for the leptoquark models of interest. We emphasize that the corrections to g-2 rising from all these models are negative, and therefore they cannot accommodate the muon magnetic moment deviation, but $1\sigma$ bounds can still be placed under the assumption the anomaly is otherwise resolved. The overall $\tilde{R}_2$ contribution is quite small and hence no meaningful limit could be placed.} 
\label{g2table}
\end{table}

\begin{figure*}[!t]
\centering
\includegraphics[scale=1]{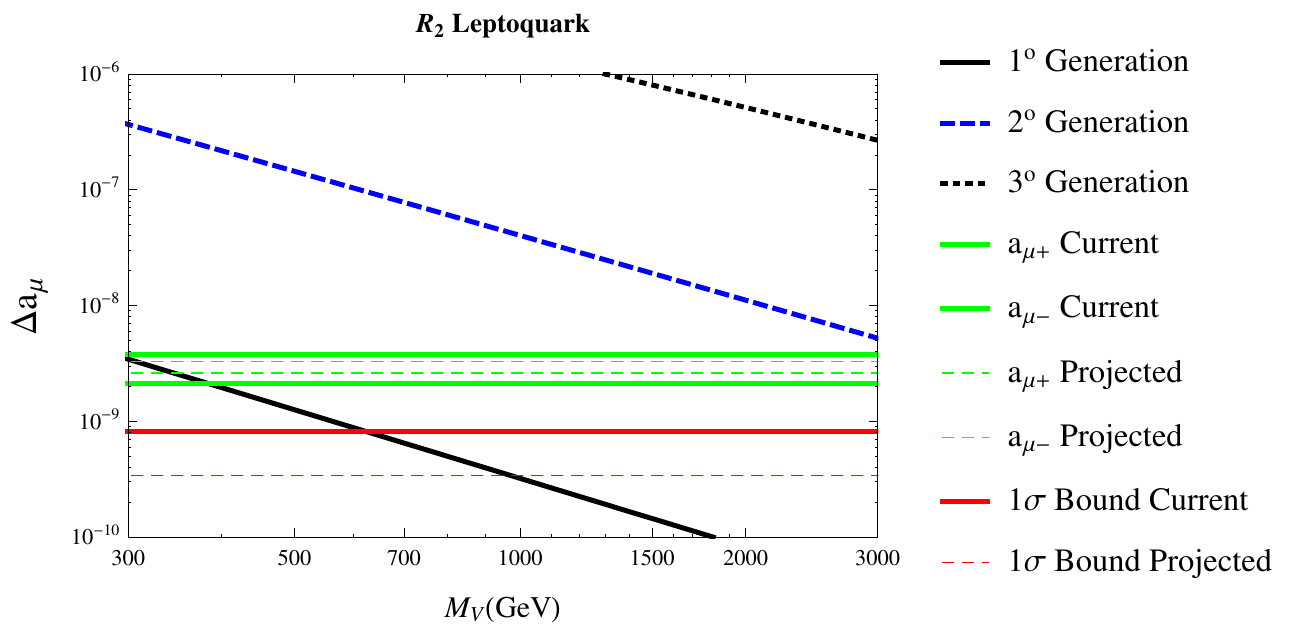}
\caption{Contributions to the muon magnetic moment coming from first,second and third generation leptoquarks $R_2$ for unity leptoquark couplings. All individual contributions scale as the square of the couplings and are negative. We plotted the absolute value of the contributions compared to current measurements. The green solid (dashed) horizontal lines are the current and projected $\Delta a_{\mu}$ . The red ones are the current (projected) $1 \sigma$ limits assuming the anomaly is otherwise resolved.}
\label{gmu}
\end{figure*}

In Fig.\ref{gmu} we show contributions to the muon magnetic moment for the LQ $R_2$ with couplings of order one. The green solid (dashed) horizontal lines are the current and projected $\Delta a_{\mu}$. The red ones are the current (projected) $1\sigma$ limits assuming the anomaly is otherwise resolved. One can clearly notice that the quark mass running in the loop is important to the overall correction to g-2 since the different generations yield different results. The contributions to the muon magnetic moment are negative for all generations. Anyhow, we can still place limits on the masses of the LQs since their contributions have to be within the $1\sigma$ error bars. That being said we derive the limits: $M_V >600$~GeV (first generation), $M_V >8$~TeV (second generation), $M_V >50 $~TeV (third generation). We point out that our limits are valid for couplings of order one. Since $\Delta a_{\mu} \propto {\rm coupling^2}$ the limits are strongly sensitive to the couplings strength.

Focusing on $650$~GeV $R_2$, we find the following constraints on couplings:  $\lambda \leq 1$ (first generation); $\lambda \leq 0.1$ (second generation); $\lambda \leq 0.01$ (third generation), where $\lambda's$ refer to the muon-leptoquark couplings. A summary of those limits is presented in Table \ref{g2table}. In conclusion, we emphasize that a first generation $650$~GeV $R_2$ LQ is not ruled out by g-2 and it is in principle a suitable candidate to explain the CMS excess. 

As for the $\tilde{R}_2$ LQ, the overall corrections to g-2 are negative and small enough so that no meaningful limit can be placed either on the couplings or the masses. Thus, $\tilde{R}_2$ $650$~GeV leptoquarks are also viable candidates to the explain the CMS excess. 

\subsection{$K$ and $B$-physics Constraints}
Scalar LQs such as the ones presented in Eq.~(\ref{lepto_lag1}) contribute to the $K$ and $B$ meson mixing, as well as leptonic and semi-leptonic meson decays from those mesons. Therefore, using current data important constraints might be derived. We refer to Ref.\cite{Saha:2010vw} for limits on various LQs models. Here we are focused on the LQ $\tilde{R_2}$ because it the relevant one for our further discussions. In this context of mesons mixing, meson oscillations $M^0-\overline{M^0}$ can have a new LQ mediated amplitude, with i-type leptons and some scalar LQ. The amplitude is proportional to $(\lambda_{ik}\lambda_{i})^2$, that rises from the effective operator $[\bar{b} \gamma_{\mu} P_R d][\bar{d} \gamma_{\mu} P_R b]$. As for the leptonic $(B_{d,s}^0 \rightarrow l^+l^-, K^0 \rightarrow l^+l^-)$ and semi-leptonic $(b \rightarrow d l^+ l^-, s \rightarrow dl^+l^-)$ decays, they are also sensitive to the aforementioned product of couplings. For the LQ $\tilde{R_2}$, the relevant interactions are the four-fermion operators $G(\overline{d}_R \nu_L)(\overline{\nu}_L d_R)$ and $G(\overline{d}_R e_L)(\overline{e}_L d_R)$, where $G \sim \lambda^2/m^2_{LQ}$. In the Tables 2-4, we summarize the updated the bounds on the product of lambdas for a $650$ GeV LQ following the recipe of Ref.\cite{Saha:2010vw}.
In the first column of Tables II-IV we exhibit bounds on a variety of coupling products. Second and third column are the bounds on the real and imaginary component of the products rising from the mixing study. In third column we show which decay channel has been studied, whereas in the fourth the respective constraint.
\begin{table} [htbp]
\begin{center}
\begin{tabular}{||c|c|c|c|c||}
\hline 
{ Yukawa}  &\multicolumn {4}{|c||}{{Bounds}}\\ 
\cline{2-5} 
 $\lambda_{(\cdot \cdot)} \lambda_{(\cdot \cdot)}^{\ast}$ & \multicolumn {2}{|c|}{{ From Mixing}}&\multicolumn {2}{|c||}{{ From Decay}}\\
\cline{2-5}
  & {Real part}&{Imag. part}& {Channel} & {Bound} \\
\hline 
\hline
$(11)(12)^*$&$1.2 \times 10^{-3}$&$1.2 \times 10^{-3}$&${K}^+ \rightarrow \pi^+ e^+ e^-$&$  2.6 \times 10^{-4}$\\
 & & &${K}_L \rightarrow \pi^0 e^+ e^-$&$ 6.0 \times 10^{-6}$\\
\hline 
$(21)(22)^*$&$1.2 \times 10^{-3}$&$1.2 \times 10^{-3}$&${K}^+ \rightarrow \pi^+ \mu^+ \mu^-$&$ 2.0 \times 10^{-4}$\\
 & & &${K}_L \rightarrow  \mu^+ \mu^-$&$ 1.4 \times 10^{-6}$\\
\hline
$(31)(32)^*$ &$ 1.2 \times 10^{-3}$&$ 1.2 \times 10^{-3}$&
$K_L\to\pi^0\nu\bar\nu$ & $ 1.0 \times 10^{-4}$\\
\hline
\end{tabular}
\end{center}
\caption{Bounds coming from $K^0 - \overline{K}^0$ mixing and correlated decays.} 
\label{tab:kk}
\end{table}
%
\begin{table} [htbp]
\begin{center}
\begin{tabular}{||c|c|c|c|c||}
\hline 
{ Yukawa}  &\multicolumn {4}{|c||}{{Bounds}}\\ 
\cline{2-5} 
$\lambda_{(\cdot \cdot)} \lambda_{(\cdot \cdot)}^{\ast}$  & \multicolumn {2}{|c|}{{ From Mixing}}&\multicolumn {2}{|c||}{{ From Decay}}\\
\cline{2-5}
  & {Real part}&{Imag. part}& {Channel} & {Bound} \\
\hline 
\hline
 $(11)(13)^\ast$  & $0.03$ & $0.03$ & $B^+ \rightarrow \pi^+ e^+ e^-$ & $ 1.1 \times 10^{-4}$\\
\hline
$(21)(23)^\ast$ & $0.03$ & $0.03$ & $B^+ \rightarrow \pi^+ \mu^+ \mu^-$&$ 1.0 \times 10^{-4}$\\
\hline
 $(31)(33)^\ast$ & $0.03$ & $0.03$ &
$B^+\rightarrow\pi^+\nu\bar\nu$ & $4.3\times 10^{-3}$\\
\hline
\end{tabular}
\end{center}
\caption{Bounds coming from $B_d - \overline{B_d}$ mixing and correlated decays.} 
\label{tab:kk}
\end{table}
\begin{table} [htbp]
\begin{center}
\begin{tabular}{||c|c|c|c|c||}
\hline 
{ Yukawa}  &\multicolumn {4}{|c||}{{Bounds}}\\ 
\cline{2-5} 
$\lambda_{(\cdot \cdot)} \lambda_{(\cdot \cdot)}^{\ast}$  & \multicolumn {2}{|c|}{{ From Mixing}}&\multicolumn {2}{|c||}{{ From Decay}}\\
\cline{2-5}
  & {Real part}&{Imag. part}& {Channel} & {Bound} \\
\hline 
\hline
$(12)(13)^*$ & $0.02$ & $0.02$ & $B_d \rightarrow K^0 e^+ e^-$ & $ 7.7 \times 10^{-5}$\\
\hline
$(22)(23)^*$ & $0.02$ & $0.02$ & $B_d \rightarrow K^* \mu^+ \mu^-$ & $ 2.3 \times 10^{-4}$\\
\hline
$(32)(33)^*$ & $0.02$ & $0.02$ &
$B^+\rightarrow K^+\nu\bar\nu$ & $ 2.0 \times 10^{-3}$\\
\hline
\end{tabular}
\end{center}
\caption{Bounds coming from $B_s - \overline{B_s}$ mixing and correlated decays.} 
\label{tab:kk}
\end{table}
Notice that these limits are complementary to the muon magnetic moment one, but are mostly sensitive to the non-diagonal couplings. We can conclude that order of one diagonal couplings are consistent with data as long as suppressed non-diagonal couplings 
are used. In other words, a $550-650$~GeV LQ might address the muon magnetic moment while being consistent with $K$ and $B$ physics limits.

We note that other interesting correlations for leptoquark models exist, with the observable $R_K$, which is the ratio of the branching fractions $\overline {B} \rightarrow \overline{K} \mu \mu $ and $\overline {B} \rightarrow \overline{K} e e $, and for which the recent LHCb measurement shows a $2.6 \sigma$ deviation. We refer to \cite{Hiller:2014yaa} for more details.

 After discussing proton decay, muon magnetic moment and electroweak constraints, we summarize in the next section the collider bounds.
\subsection{Collider Bounds}
After LQs are pair produced through QCD interactions, they decay to final states containing jets and leptons. The bounds on the LQ masses thus depend on the branching fraction to the various channels. Searches have been conducted for all the generations. 
CMS has studied first generation LQs with 19.6 fb$^{-1}$ of data at the 8 TeV LHC. Events with $eejj$ and $e \nu jj$ final states have been targeted. For the $eejj$ channel, events with exactly two electrons with $p_T > 45$ GeV, and at least two jets with $p_T > (120, 45)$ GeV have been selected; subsequently, for different LQ masses, the invariant mass $m_{ee}$ of the electrons, the scalar sum $S_T$  of the $p_T$ of the two leading jets and the two electrons, and the average electron-jet invariant mass $m_{ej}$, obtained from the two electrons and two jets, were optimized. For the benchmark point of $m = 650$ GeV, the cuts were $m_{ee} > 155$ GeV, $S_T > 850$ GeV, and $m_{ej} > 360$ GeV.  36 signal events were observed, against an expected background of $20.49 \pm 2.14 \pm 2.45 (syst)$, yielding a significance of $2.4\sigma$. Interestingly, the number of signal events is lesser than the expected signal for a LQ decaying purely into electrons and jets with $100 \%$ branching ratio. This implies that there is non-zero branching to other states. In the $e \nu jj$ channel, the optimized cuts are $S_{\rm T} > 1040$~GeV, $E^{\rm miss}_{\rm T} > 145$~GeV, $m_{\rm ej} > 555$~GeV and  $m_{\rm T, e\nu} > 270$~GeV. 18 signal events were observed over a background of $7.54\pm 1.20\pm 1.07 ({\rm syst})$, giving a significance of $2.6$.
The CMS study cannot exclude LQs with masses of $650$ GeV, and branching ratio $\beta$ to an electron and a quark $\beta < 0.15$. On the other hand, LQs with $\beta = 1 \, (0.5)$ are ruled out up to masses of $950 \, (845) GeV$. From the above discussion of bounds, it is clear that if the signal in the $eejj$ and $e \nu jj$ channels are real, LQs have significant branching to other states. Due to the stringent constraints on branching to the second generation, it is natural to consider a scenario where the LQ decays mainly into dark matter. In the next section, we propose models where the latter is fulfilled.
\section{Leptoquarks and Dark Matter}
Given the nature of the signal seen by CMS, it is clear that considering models where the LQ decays primarily to dark matter and jets is well-motivated. In this Section, we will provide the simplest such models. As we shall see, the confluence of proton decay, dark matter direct and indirect detection bounds, and Higgs invisible decay constraints leads to a few very limited and predictive scenarios.

\subsection{The Model}

In this work we focus on the LQ $\widetilde{R}_2 = (3,2,1/6)$, to avoid tree-level proton decay, as described previously. It is natural to connect this field to the left-handed quark doublet. To ensure the stability of the dark matter candidate, we impose a $Z_2$ symmetry under which the dark sector is odd, whereas the SM + LQ sector is even. This requires at least two fields ($S$ and $\chi$) in the dark sector. Then, to satisfy $SU(2)_L$ charges, an interesting possibility is to also have a gauge singlet scalar $S$ and a color-neutral $SU(2)$ triplet fermion $\chi$. Thus, in addition to the SM, our model has the following content:
\bea
\widetilde{R}_2 &=& (3,2,1/6)_{+} \nonumber \\
\chi  &=&  (1,3,0)_{-} \\
S &=& (1,1,0)_{-} \,\,,
\eea
where we have also displayed the charge under the $Z_2$. Notice that to conserve lepton number, $\chi$ has lepton number $L = 1$. We will require that $\chi$ has a Dirac mass term \footnote{We could also accommodate a Majorana mass term if it is generated at a high scale. This won't change our results much.}.
Several variations of the above model, with different $SU(2)_L$ charge assignments for $S$ and $\chi$, are possible. These options are outlined in Section \ref{variations}. Here, we will consider mainly the model with the $SU(2)_L$ triplet fermion and singlet scalar. 
There are no four dimensional operators connecting the SM + LQ sector to the dark sector ( apart from several ``Higgs-portal terms" for $S$, which we will display soon). Up to dimension five, the operators we can write down are the following:
\be \label{modelL1}
L \, \supset \, L_{SM} \, - \lambda_d^{ij} \bar{d}_{R}^i \widetilde{R}_2^T\epsilon L_L^j  \,- \, \frac{1}{\Lambda_1} h_i S \overline{Q} \chi \widetilde{R}_2 \,- \, \frac{1}{\Lambda_2} h^{\prime}_i S \overline{L} \chi H\,\, .
\ee
We have not displayed the kinetic and mass terms for the new fields. 
The scales $\Lambda_1$ and $\Lambda_2$ depend on the scale at which new physics sets in
(the corresponding effective operators can be realised within renormalizable mediator models),
and will be taken to be  $\sim 1$ TeV.

We now turn to a discussion of the bounds on the Yukawa parameters $\lambda_{ij}$ in Eq.~\ref{modelL1}.
\subsection{Bounds on Yukawas $\lambda_{ij}$}
The approximate constraint coming from collider processes is 
\be \label{colliderunambiguous}
{\rm LHC:} \,\,\,\,\,\, \lambda_{ij} \, \gsim \, 10^{-8} \,\,,
\ee
which comes from ensuring decay at the collision point. This bound is weak enough that it will not affect any of our subsequent analysis. Moreover, the LQ does not couple to the top quark, so there is no $m_t$ enhancement to $\mu \rightarrow e \gamma$ due to a charged lepton chirality flip at one loop. Thus, bounds on the Yukawas coming from flavor violation in the charged lepton sector are expected to be small for this model. Much stronger bounds rise from neutral meson mixing and semi-leptonic and leptonic decays though. These bounds mostly affect off-diagonal entries of the Yukawa matrix as discussed in the Section 2.3. 
\subsection{Spectrum and Decays}

It is useful to write
\be
\chi \, = \ \left( \begin{array}{cc}
\frac{1}{\sqrt{2}}\chi^0 & \chi^{+} \\
\chi^{-} & - \frac{1}{\sqrt{2}}\chi^0
 \end{array} \right) \,\,.
\ee
One obtains the same tree level mass for the neutral and charged eigenstates, while there is a loop level splitting given by loops of SM gauge bosons. The mass splitting of two states with charges $Q$ and $Q^{\prime}$ is given by \cite{Cirelli:2009uv}
\bea  
& &\Delta M = M_Q -  M_{Q'} = \frac{\alpha_2 m_{\chi}}{4\pi}\left\{(Q^2-Q^{\prime 2})s_{\rm W}^2 f(\frac{M_Z}{m_{\chi}}) \right. \nonumber\\
& & \left. +(Q-Q')(Q+Q'-2Y)
\bigg[f(\frac{M_W}{m_{\chi}})-f(\frac{M_Z}{m_{\chi}})\bigg]\right\}
\eea
where
\be
f(r) =\left\{\begin{array}{ll}
+r  \left[2 r^3\ln r -2 r+(r^2-4)^{1/2} (r^2+2) \ln A\right]/2 \\
-r \left[2r^3\ln r-k  r+(r^2-4)^{3/2} \ln A\right] /4
  \end{array}\right.
\ee where the upper and lower results are for a fermion and scalar respectively, with $A = (r^2 -2 - r\sqrt{r^2-4})/2$ and $s_{\rm W}$ the sine of the weak angle.

For $m_{\chi} \sim \mathcal{O}(100-500)$ GeV, one obtains $\Delta M \, \sim \, 166$ MeV \footnote{We note that there are corrections to this mass splitting coming from the dimension 5 operators in Eq.~\ref{modelL1}. However, these corrections will not significantly change our results.}.
The decay widths of the LQ are given by 
\bea
\Gamma_{\widetilde{R}_2 \rightarrow (lj)} & \simeq & \frac{\lambda^2 N_f m_{\widetilde{R}_2}}{4\pi} \\
\Gamma_{\widetilde{R}_2 \rightarrow (\chi j)} &\simeq & \frac{h^2 N_f}{8 \pi^2} \frac{m^3_{\widetilde{R}_2}}{\Lambda^2_1},
\eea where $N_f$ is the number of colors of the quark in the final state.
We have denoted a leptonic final state by $l$, and a quark final state with $j$. As stated in Ref.\cite{Bai:2014xba}, a $\chi^2$ fit to the data prefers a $550$~GeV LQ with $BR_{\nu j,lj}\simeq 0.15$, or a $650$~GeV LQ with $BR_{\nu j} \simeq 0.13$ and $BR_{l j} \simeq 0.21$. Our model is suitable for a $550$~GeV LQ since the leptonic branching ratios are the same. We point out that mild changes are expected in the aforementioned meson related bounds in case one targets a $650$~GeV LQ. In order to fulfil the former setup, we need $h/\lambda \simeq 5.6$ for $\Lambda=1$~TeV or $h/\lambda \simeq 56$ for $\Lambda=10$~TeV, which is feasible. We note that the production of LQs at the LHC depends on the strong coupling constant $\alpha_s$, and it is nearly independent of the couplings $\lambda_{1i}$. In fact, the CMS study took a value of $\lambda_{1i} = 0.3$, which is small enough so that single LQ production can be neglected. In our model, this translates to $h \sim \mathcal{O}(1)$, which is natural. Since the excess is mild the quoted numbers are very sensitive to further investigation from the collaboration, but it is clear we have a plausible model.
We note a small caveat. After electroweak symmetry breaking, the term $\frac{1}{\Lambda_2} h^{\prime}_i S \overline{L}\chi H$ induces the decay $\chi \, \rightarrow l S$, with a decay width $\Gamma \sim (v_{ew}/\Lambda_2)^2 m_{\chi}$. This will induce $\widetilde{R}_2 \rightarrow (l + j + \met)$. However, we see later that  dark matter constraints will force us into a benchmark scenario where 
we would not expect this scenario to have survived the jet and electron $p_T$ cuts in the CMS $e \nu jj$ study \footnote{In principle, this signal can also be avoided by choosing $\Lambda_2$ large enough that the $\chi$ decay happens outside the detector. }.
\section{Dark Matter Phenomenology}
In this section, we discuss the dark matter phenomenology of our model. There are two possibilities: either one of $\chi$ and $S$ can be the dark matter candidate. We first discuss the case of scalar dark matter $S$, and subsequently discuss the bounds on the fermionic candidate $\chi$. The dark matter stability is guaranteed by the discrete symmetry. We note that in the absence of the symmetry, there is no need for the singlet scalar $S$, and the fermion $\chi$ is enough to satisfy the quantum numbers. However, there is a fast decay channel through the $ h^{\prime}_i  \overline{L} \chi H$ term in that case.
\subsection{Scalar Dark Matter}
\begin{figure*}[!t]
\mbox{\includegraphics[scale=0.6]{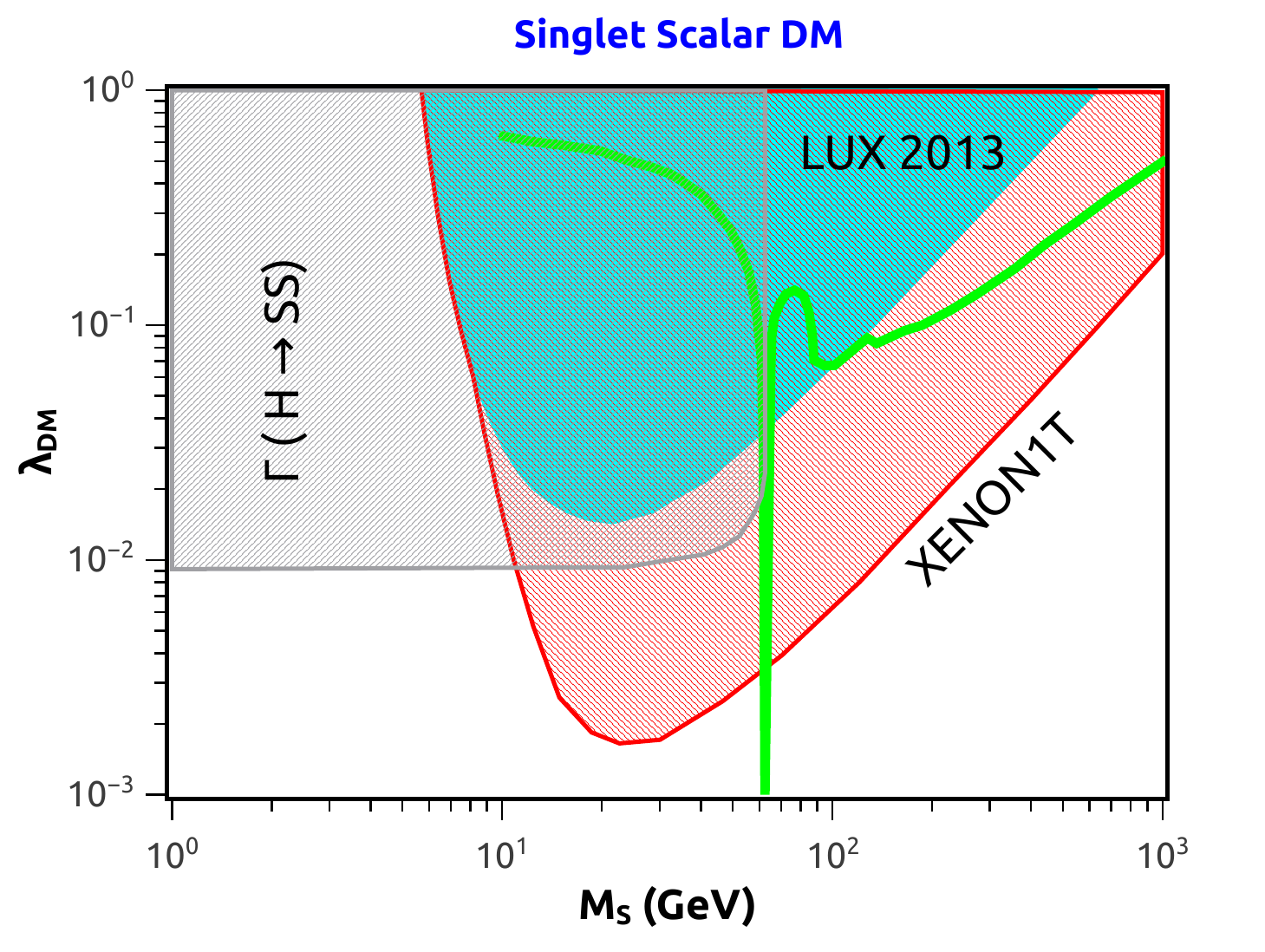}\includegraphics[scale=0.7]{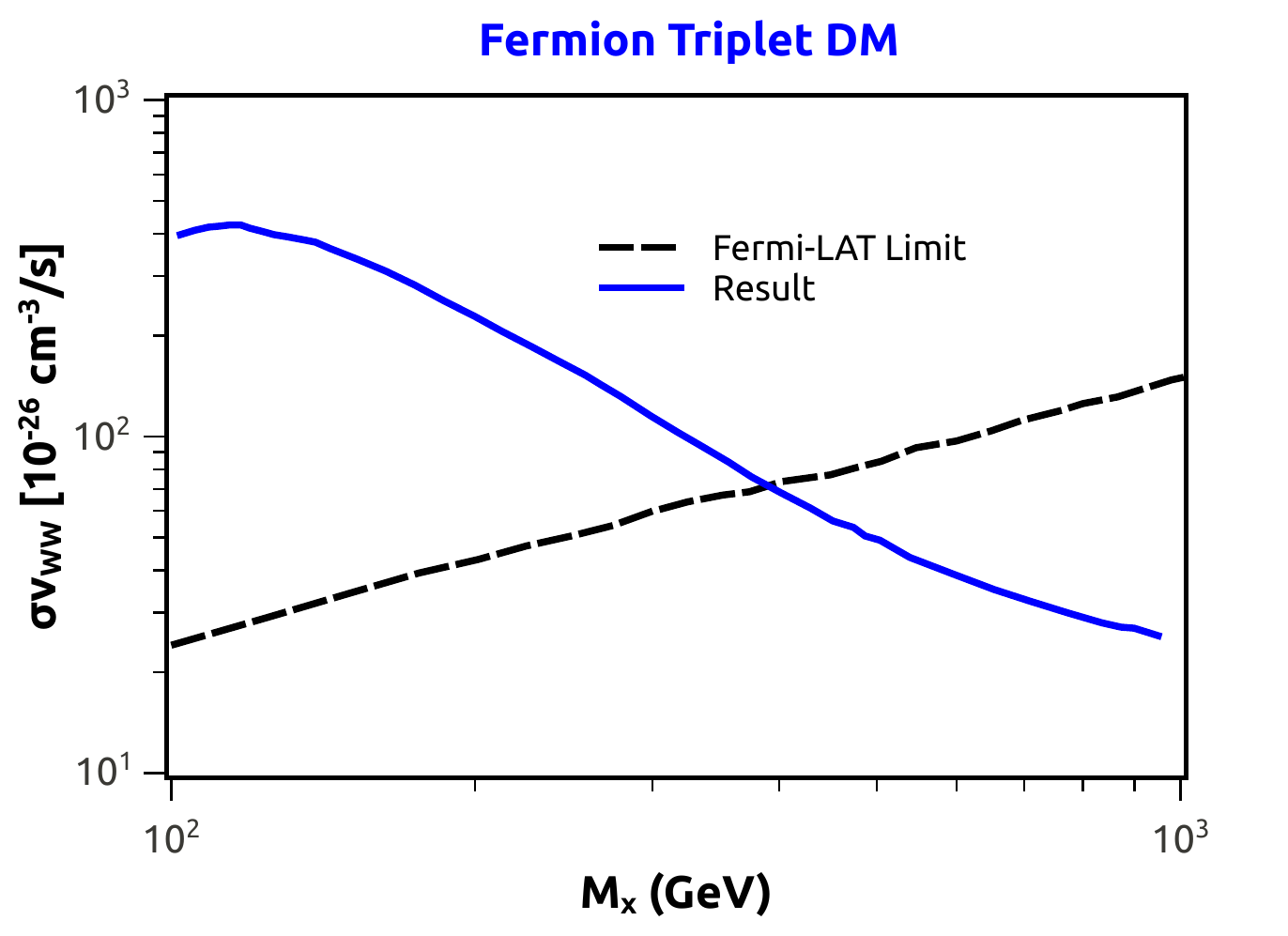}}
\caption{ {\it Left}: Direct detection and Higgs width constraints on the singlet scalar dark matter scenario. Current LUX bound excludes masses below $\sim 100$~GeV. Projected XENON1T limit will be able to rule out the entire few $GeV-1$~TeV mass range. {\it Right:} Annihilation cross section into $WW$ final states for the triplet fermion dark matter model along with the Fermi-LAT dwarf galaxy exclusion limit. We conclude that Fermi-LAT rules out fermion triplet dark matter below $\sim 400$ GeV.}
\label{Graph1}
\end{figure*}

The scalar $S$ can annihilate through the Higgs portal. Specifically, the relevant terms in the Lagrangian are
\be
L \, \supset \, \frac{1}{2} \partial_{\mu} S \partial^{\mu} S - \frac{m^2_0}{2}S^2 - \frac{\lambda_s}{4}S^4 - \frac{\lambda_{DM}}{4} S^2 H^{\dagger} H \,\,.
\ee
After electroweak symmetry breaking, shifting the Higgs by its vacuum value $h \rightarrow h + v_{ew}$ (where $h$ denotes the physical Higgs), and using $H^{\dagger}H = \sqrt{2}h v_{ew}$, the $S-$dependent part of the potential becomes
\be \label{Vfinal}
V \, = \, \frac{1}{2}(m^2_0 + \frac{\lambda_{DM}}{4} v^2_{ew})S^2 + \frac{\lambda_s}{4}S^4 + \frac{\lambda_{DM}}{2\sqrt{2}} v_{ew} S^2 h + \frac{\lambda_{DM}}{2}S^2 h^2 \,\,.
\ee
The mass of the dark matter is given by $m^2_S = m^2_0 + \frac{\lambda_{DM}}{4}v^2_{ew}$. We will assume that the couplings satisfy conditions for the existence of a stable vacuum, and a thermal history for $S$ (typically, this amounts to requiring $\lambda_{DM} \geq 10^{-8}$).
The coupling $\lambda_{DM}$ sets the relic abundance as well as the spin independent scattering cross section of the dark matter candidate. There are only two free parameters: $\lambda_{DM}$ and the mass $m_S$.
If $S$ is lighter than half of the Higgs mass, the  $h \rightarrow SS$ channel becomes kinematically available and changes the measured invisible width of Higgs. The current limit on this is roughly $10-15\%$ \cite{Belanger:2013xza,Ellis:2013lra}, and a projected bound of $5\%$ at $14$~TeV LHC after $300 {\rm fb^{-1}}$ has been claimed \cite{Peskin:2012we}. We use the latter. The Higgs branching ratio into $S$ is
\begin{equation}
{\rm Br}_{h \rightarrow SS} = \frac{\Gamma_{h \rightarrow SS}}{\Gamma_{vis}+\Gamma_{h \rightarrow SS}}
\end{equation}where $\Gamma_{vis} = 4.07$~MeV for 
$M_h =125$~GeV and,
\begin{equation}
\Gamma_{h \rightarrow SS} = \frac{\lambda^2_{DM} v^2_{ew}}{8 \cdot 32 \pi M_h}\left(1-\frac{4 m^2_S}{M^2_h}\right) 
\end{equation}
This bound is depicted in Fig.\ref{Graph1} (grey shaded region). Our results agree well with Refs.\cite{deSimone:2014pda,Farina:2013mla}.
It is clear that $\lambda_{DM}$ has to be smaller than $\sim 10^{-2}$ to obey the Higgs invisible width limit. 
The relic abundance and the scattering cross section of the scalar as a function of the two relevant parameters $\lambda_{DM}$ and the mass $m_S$ has been computed. The relic density is driven by the s-channel annihilation into Standard Model particles. There is a subdominant annihilation into $hh$, through the quartic scalar interaction $S^2 h^2$ in Eq.\ref{Vfinal} and by t-channel $S$ exchange. In Fig.\ref{Graph1} the region determined by the green curve yields the right abundance in agreement with Ref.\cite{Queiroz:2014yna,deSimone:2014pda}. The cyan region is ruled out by LUX bounds based on the 2013 data \cite{LUXbound}, whereas the red one is excluded by the projected XENON1T results \cite{Cushman:2013zza}. Our numerical calculations are performed using micrOMEGAS \cite{micromegas}. 
Clearly, thermally produced singlets $S$ are excluded by direct detection data for $m_S \lesssim 100$~GeV, unless it is at the Higgs resonance. In order for the LQ to decay to dark matter, we need $m_S \, \leq \, 325$ GeV. Thus from Fig.\ref{Graph1} we see that the next run of XENON1T and LUX will decisively rule out this model.
\subsection{Fermionic Triplet Dark Matter}
The field $\chi$ is a fermionic triplet with a Dirac mass term. Its couplings to the Standard Model relevant for relic density calculations and indirect/direct detection bounds is that of supersymmetric Wino dark matter \footnote{Winos are Majorana, whereas $\chi$ is Dirac. However, for the purposes of indirect/direct bounds in this case, the results are similar.}.
Triplet dark matter is under severe tension in the mass range we are considering (the case of the Wino has been studied in detail \cite{winobounds}). We outline the main bounds here. 
For the triplet fermionic dark matter $(\chi^0, \chi^{\pm})$, the relevant portion of the interaction Lagrangian with the Standard Model comes through gauge interactions:
\begin{eqnarray}
{\cal L}_{\rm int} &=&
  -\frac{e}{s_W}
  \left(
    \overline{\tilde{\chi}^0}\gamma^\mu\tilde{\chi}^-W^\dagger_\mu
    +
    h.c.
  \right)\nonumber\\ 
& &  +
  e \frac{c_W}{s_W}\overline{\tilde{\chi}^-}\gamma^\mu\tilde{\chi}^-Z_\mu
  +
  e\overline{\tilde{\chi}^-}\gamma^\mu\tilde{\chi}^-A_\mu \ .
\end{eqnarray}
Here, $e$ is the electric charge, $s_W=\sin\theta_W$ and
$c_W=\cos\theta_W$ with $\theta_W$ being the Weinberg angle. 
The leading contribution to the $\chi$-quark interaction comes from one-loop interactions. After evaluating the relevant diagrams, it turns out that
\be
\sigma^{SI} \, \sim \, 10^{-47} \, {\rm cm}^2 \,\,
\ee
in the mass range of interest. We refer to \cite{Hisano:2010fy} for further details. This may be probed by XENON1T.
On the other hand, the model is severely constrained by indirect detection data. Continuum photons arise from the tree level annihilation process $\chi^0 \chi^0 \rightarrow W^{+} W^{-}$. Fermi-LAT dwarf galaxy data \cite{fermi} rules out triplet dark matter masses up to around $400$ GeV, while galactic center data rules it out up to around $700$ GeV for either NFW or Einasto profiles \cite{Hryczuk:2014hpa}. Such bounds become stronger (weaker) if one considers steeper (less steep) profiles, such as those possibly motivated by massive black holes \cite{Gonzalez-Morales:2014eaa} (isothermal models).

\subsection{Variations and Other Options for Dark Matter} \label{variations}
In this subsection, we briefly discuss other dark matter possibilities and variations of the model we described. We will only consider dark sectors with $Y = 0$. Candidates with $Y \neq 0$ have vector-like interactions with the $Z$ boson, that gives spin-independent scattering cross-sections which are ruled out by current direct detection constraints. This conclusion could be altered if for example $\chi$ dark matter is Majorana, in which case it cannot have a vector-like coupling to $Z$. This is reminiscent of Higgsino dark matter in supersymmetry. However, such a mass term would violate lepton number in our case, and would have to be generated at a high scale, as mentioned before. We are thus led to the variations outlined in this Section.
\subsubsection{Fermion Singlet Dark Matter}
A simple variation of the model presented before is
\bea
\widetilde{R}_2 &=& (3,2,1/6)_{+} \nonumber \\
\chi  &=&  (1,1,0)_{-} \\
S &=& (1,1,0)_{-} \,\,,
\eea
with a Lagrangian still given by Eq.~\ref{modelL1}. As we have mentioned before, it is possible in this variation to accommodate a gauge singlet Dirac fermion dark matter, if $\chi$ is lighter than $S$. 
The dark matter annihilation and scattering cross sections are set by the fermionic dark matter Higgs portal \footnote{The $Z_2$ symmetry forbids a term like $\overline{\chi} \chi S$. Such a term could have been allowed if only $\chi$ was odd under the $Z_2$ symmetry, while $S$ was even. In that case, a term like $S H^{\dagger}H$, allowed by the symmetries, would have mixed $S$ with the Higgs, allowing for a Higgs portal connection for $\chi$ through the $\overline{\chi} \chi S$ term. Such models have been studied by \cite{Esch:2013rta}. In our case, this charge assignment under $Z_2$ would forbid the all-important $\frac{1}{\Lambda_1} h_i S \chi \overline{Q} \widetilde{R}_2$ term.}. This has been studied in detail by \cite{Fedderke:2014wda}. In this context the coupling between $\chi$ and $H$ can be written as
\be
L \, \supset \, \frac{1}{\Lambda}(v_{ew}h + (1/2)h^2)(\cos{\xi} \overline{\chi}\chi + \sin{\xi} \overline{\chi} i\gamma_5 \chi)\,\,,
\ee
where $\cos{\xi}$ and $\sin{\xi}$ control the relative strengths of the $CP-$conserving and violating terms. Similar to the scalar Higgs portal, direct detection experiments severely constrain this model. Current LUX bounds can accommodate $m_{\chi} \sim 300$ GeV for $\cos{\xi}^2 \sim 0.5$. However, for smaller dark matter masses, the pseudoscalar component must dominate rapidly to evade bounds. This is an unnatural tuning of the Lagrangian. 
The resonant Higgs portal with $m_{\chi} \sim 60$ GeV is allowed for all values of $\cos{\xi}$. However, this is a very specific dark matter mass that is not a priori motivated from UV physics. 
In the next section we discuss another possible variation of the dark sector.
\subsubsection{Scalar Triplet Dark Matter}
\begin{figure}[!htp] \label{Graphscalartriplet}
\begin{center}
\includegraphics[width=8cm]{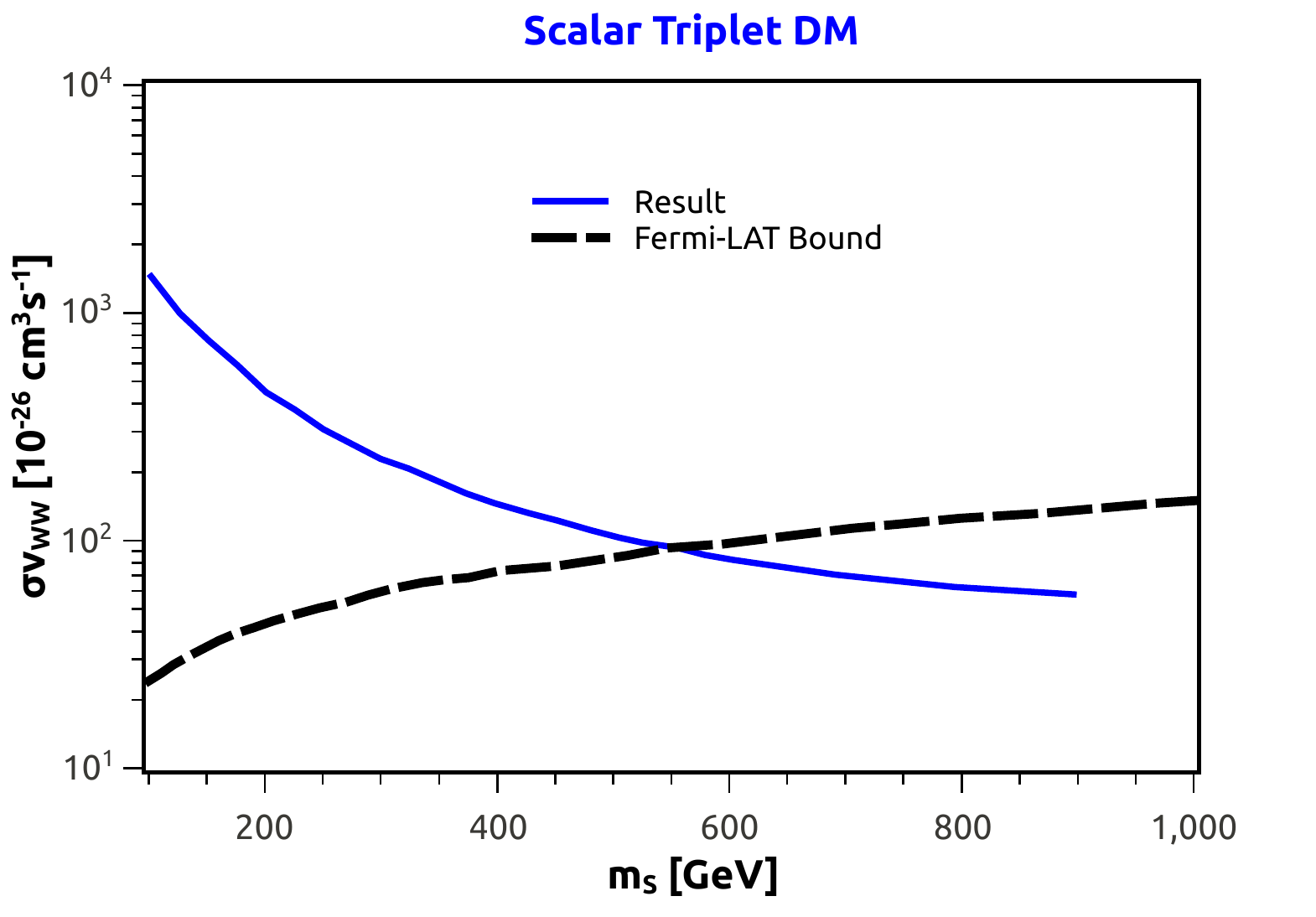}
\caption{The viable and excluded mass range of the scalar triplet model. The blue line is the annihilation cross section to $WW$ final state in the current Universe. The black line is the bound from Fermi-LAT data.}
\end{center} 
\end{figure}

Another variation of the original model that may be considered is to switch the $SU(2)$ charges of the scalar and fermion, leading to the following particle content
\bea
\widetilde{R}_2 &=& (3,2,1/6)_{+} \nonumber \\
\chi  &=&  (1,1,0)_{-} \\
S &=& (1,3,0)_{-} \,\,,
\eea
with the same Lagrangian in Eq.~\ref{modelL1} as before. A quartic interaction with the Higgs is also possible, which will induce mass splitting within the multiplet. When the interaction strength is small, the splitting reverts to the usual $166$ MeV induced by gauge couplings.
The constraints from indirect detection are very stringent in the mass range we are considering \cite{minimaldarkmatterpapers} as once can see in Figure \ref{Graphscalartriplet}, where we display the annihilation cross section of scalar triplets into $WW$ final states and the Fermi bound. It is evident that Fermi dwarf galaxy data rules out scalar triplet dark matter up to $\sim 550$ GeV. 
\subsection{Comments on Collider Signatures}
In the previous sections, we have described the dark matter candidates that are capable of coupling to LQs and satisfying all constraints from colliders, proton decay, and direct and indirect detection experiments. The possible candidates are quite limited:

\begin{enumerate}

\item A scalar singlet in the mass range of $100$ GeV $- \, 325$ GeV, coupled to the Higgs portal, evades LUX bounds; 

\item A fermion singlet in the mass range of $300$ GeV with a significant pseudoscalar coupling;

\item A scalar or fermionic singlet in the resonant Higgs portal, with mass $\sim 60-65$ GeV.
\end{enumerate}
Scalar and fermionic triplets, which are the other possibilities, are ruled out by indirect detection in the mass range of interest.
We now briefly discuss the opportunities for collider studies of the scenarios above.  After the LQ with mass $\sim 650$ GeV decays to $\chi$ and $S$, there is a further decay of $\chi$ to a lepton and $S$. Thus, the final state from each LQ decay consists of $l, j$ and $\met$. It would be very interesting to probe this scenario in $2l \, + \, 2j \, + \, \met$ events. For dark matter in the mass range $100-325$ GeV that we are considering in scenarios $(1)$ and $(2)$, the lepton and jets are expected to be significantly softer than the cuts employed in the LQ $e\nu jj$ study. Moreover, since $m_{LQ} \, \sim \, m_{S} + m_{\chi}$, this is a compressed scenario where the missing energy release is small. It would be very interesting to probe this scenario in events with significant boosting due to the presence of high $p_T$ initial jets \cite{Dutta:2012xe}, where appreciable $\met$ to distinguish signal and background may be obtained.

\section{Conclusions}

 This paper has been motivated by the mild excess seen by CMS in the $eejj$ and $e \nu jj$ channels, in its first generation LQ search. If the excess is indeed due to a LQ in the mass range of $550-650$ GeV, then there has to be branching to other states which are not captured in those channels. Given the stringent bounds on second and third generation LQs, the idea of tying LQs to dark matter is well motivated.

Among the leptoquark models presented in Table \ref{table1}, we were left with only $R_2$ and $\widetilde{R_2}$ to avoid fast proton decay. Due to its quantum number $\widetilde{R_2}$ seems more attractive from a model building perspective. Taking $\widetilde{R_2}$ as our candidate, the gauge charges of the LQ, stability of the dark matter, and avoiding large $Z-$mediated nucleon-dark matter scattering cross-section force us to consider a limited set gauge charge assignments in the dark sector. The dark matter can be either a scalar/fermion triplet, or a scalar/fermion singlet. The former cases are ruled out by Fermi-LAT data in the mass range of interest. For the singlets, the following options emerge:
\begin{enumerate}

\item A scalar singlet in the mass range of $100$ GeV $- \, 325$ GeV, coupled to the Higgs portal; 

\item A fermion singlet in the mass range of $300$ GeV with a significant pseudoscalar coupling to the Higgs portal;

\item A scalar or fermionic singlet in the resonant Higgs portal, with mass $\sim 60-65$ GeV.
\end{enumerate}
Those finding are consistent with muon magnetic moment, proton decay, $K^0$ and $B^0$ mesons constraints. Lastly, we have also outlined collider strategies for exploring these cases. The most promising search channels would be $2l \, + \, 2j \, + \, \met$ events, with significant boosting due to the compresed nature of the spectrum.

\section{Acknowledgements}

The authors are indebted to Josh Berger for clarifying their results and Eugenio Del Nobile for important discussions concerning the muon magnetic moment.  FQ is partly supported by US Department of Energy Award SC0010107 and the Brazilian National Counsel for Technological and Scientific Development (CNPq). KS is supported by NASA Astrophysics Theory Grant NNH12ZDA001N. AS is supported by ESF grant MTT8 and
by the SF0690030s09 project and by the European Programme PITN-GA-2009-237920 (UNILHC).

\appendix

\end{document}